\documentclass[11pt]{article}

\usepackage{amssymb}
\usepackage{amsmath}
\usepackage{amsthm}
\usepackage{mathrsfs}
\usepackage{verbatim}

\numberwithin{equation}{section}

\newtheorem{thm}{Theorem}[section]
\newtheorem{prop}[thm]{Proposition}
\newtheorem{lem}[thm]{Lemma}

\newcommand{\RR}{\mathbb{R}}  
\newcommand{\LL}{\mathbb{L}}
\newcommand{\mr}{\mathring}
\newcommand{\DD}{\mathcal{D}}
\newcommand{\KK}{\mathcal{K}}
\newcommand{\ZZ}{\mathbb{Z}}
\newcommand{\Lich}{\mathrm{Lich}}
\newcommand{\mcs}{\mathcal{S}}
\newcommand{\QQ}{\mathcal{Q}}

\begin{document}

\nocite{*}

\title{A far-from-CMC existence result for the constraint equations on manifolds with ends of cylindrical type}
\author{Jeremy Leach\footnote{Supported by the NSF grant DMS-1105050; email \textbf{jleach@math.stanford.edu}} \\ Stanford University}
\date{}

\maketitle

\begin{abstract}

We extend the study of the vacuum Einstein constraint equations on manifolds with ends of cylindrical type initiated in \cite{CM} and \cite{CMP} by finding a class of solutions to the fully coupled system on such manifolds. We show that given a Yamabe positive metric $g$ which is conformally asymptotically cylindrical on each end and a 2-tensor $K$ such that $(\mathrm{tr}_g K)^2$ is bounded below away from zero and asymptotically constant, then we may find an initial data set $(\overline{g}, \overline{K})$ such that $\overline{g}$ lies in the conformal class of $g$. 

\end{abstract}


\section{Introduction}

Let $(M,g,K)$ be an initial data set where $(M,g)$ is a complete Riemannian $n$-manifold and $K$ is a symmetric 2-tensor. Throughout this paper, we will assume $n \geq 3$. The Cauchy problem in general relativity asks whether $(M,g)$ can be embedded as a spacelike submanifold of some $(n+1)$-dimensional Lorentzian manifold $(V,\hat{g})$ which satisfies Einstein's equation, and in such a way that $g$ is the metric induced by $\hat{g}$ and the second fundamental form of this embedding is given by $K$. In the absence of matter fields, this is equivalent (see, for example, \cite{C-B} or \cite{W}) to requiring that $(g,K)$ solve the system
\begin{equation} \label{eq:constr}
\begin{cases}
R(g) = 2\Lambda + |K|_g ^2 - (\mathrm{tr}_g K)^2 \\
\mathrm{div}_g K - \nabla \mathrm{tr}_g K = 0
\end{cases}
\end{equation}

\noindent where $\Lambda$ is a cosmological constant which we assume to be nonnegative. This constant will not play a central role in our analysis, though we will assume throughout that it is small with respect to  $\tau := \mathrm{tr}_g K$ in the sense that, for some positive constant $\beta_0$, we have

\begin{equation}
\beta := \dfrac{n-2}{4n} \tau^2 - \dfrac{n-2}{2(n-1)} \Lambda \geq \beta_0.
\end{equation}

\subsection{Asymptotically cylindrical and periodic ends}

Suppose the manifold $(M,g)$ has a finite number of ends $E_\ell$, each of which is of cylindrical type. This is to say that $E_\ell$ is diffeomorphic to the cylinder $\RR^+ \times N_\ell$ where $N_\ell$ is a closed $(n-1)$-manifold. We say that a metric $\check{g}_\ell$ is \textit{asymptotically cylindrical} (AC) on $E_\ell$ if $\check{g}_\ell$ and its derivatives decay exponentially to the standard product metric on $E_\ell$. More precisely, for some $\omega > 0$ we have that
\begin{equation*}
| \mr{\nabla}^k [ \check{g}_\ell - (dt^2 + \mr{g}_\ell)  ]  | = \mathcal{O}(e^{-\omega t})
\end{equation*} 

\noindent for all $k$, where $\mr{g}_\ell$ is some Riemannian metric on $N_\ell$, and $\mr{\nabla}$ is the connection associated to the cylindrical metric $dt^2 + \mr{g}_\ell$ (we may, for our purposes, weaken this condition by only requiring this decay in $m > 2$ derivatives). We say that the metric $g$ is \textit{conformally asymptotically cylindrical} if $g = w^{4/(n-2)} \check{g}$ where $\check{g}$ is AC on each end and $w$ is a positive function such that $w \rightarrow \mr{w}_\ell$ as $t \rightarrow \infty$ on $E_\ell$, where $\mr{w}_\ell$ is a smooth positive function on $N_\ell$, and that the derivatives converge at the same rate. We similarly define a slightly more general class of metrics on $M$, namely those which are asymptotically periodic (AP) on each end. Such metrics and their derivatives decay exponentially to a periodic metric $\mr{g}_\ell$ with period $T_\ell > 0$, which is the lift to $\RR \times N_\ell$ of a metric on $(\RR / T_\ell \ZZ) \times N_\ell$. A metric $\check{g}$ is \textit{conformally asymptotically periodic} if $\check{g}$ can be written as a conformal factor times an AP metric as above, but we require that the limiting function $\mr{w}_\ell$ be periodic on $E_\ell$ and have the same period $T_\ell$ as the metric $\mr{g}_\ell$. For convenience, assume that all rates of decay are $e^{-\omega t}$, and all periods are equal to $T$.

\subsection{The conformal method}

A number of approaches have been taken to solve the constraint equations under varying mean curvature assumptions (for an excellent survey, see \cite{BI}), but perhaps none has been so successful as the conformal method. The basic idea is to fix a metric $g$ on $M$ and to look for a solution $(\tilde{g},\tilde{K})$ of the constraint equations with the form 
\begin{align}
\tilde{g}_{ij} &= \tilde{\phi}^{4/(n-2)} g_{ij} \\
\tilde{K}^{ij} &= \dfrac{\tau}{n} \tilde{g}^{ij} + \tilde{\phi}^{-2 \alpha} (\sigma^{ij} + (\DD W)^{ij}).
\end{align}

\noindent Here $\alpha = (n+2)/(n-2)$, $\tilde{\phi}$ is a positive function, $\sigma$ is a given transverse traceless 2-tensor, and $\DD$ is the conformal Killing operator, which maps a vector field $X$ to a symmetric 2-tensor by

\begin{equation*}
(\DD X)^{ij} = \nabla^i X^j + \nabla^j X^i - \dfrac{2}{n} \nabla_k X^k g^{ij}.
\end{equation*}

\noindent We call any element in the nullspace of $\DD$ a \textit{conformal Killing field}. Observe that these substitutions effectively replace the unknowns $(g,K)$ in the constraint equations with new unknowns $(\tilde{\phi},W)$, and (see, for example, \cite{G}) the task of solving the constraint equations is then reduced to solving the coupled system

\begin{equation} \label{eq:lcby}
\begin{cases}
\Delta_g \tilde{\phi} - c_n R_g \tilde{\phi} - \beta \tilde{\phi}^\alpha + c_n |\sigma + \DD W|_g ^2 \tilde{\phi}^{-\gamma} = 0 \\
\Delta_\LL W = \dfrac{n-1}{n} \tilde{\phi}^{\gamma - 1} \nabla \tau
\end{cases}
\end{equation}

\noindent where 

\begin{equation*}
c_n = \dfrac{n-2}{4(n-1)}, \hspace{.5cm} \gamma = \dfrac{3n-2}{n-2}
\end{equation*}

\noindent and $\Delta_\LL := -\mathrm{div}_g \DD$. With the assumptions above, we seek a solution to the system (\ref{eq:lcby}) with the hypothesis that on each end $|\sigma|_g ^2$ converges to some function $\mr{\sigma}_\ell ^2$ which is not identically zero. Assume that each $\mr{\sigma}_\ell ^2$ is independent of $x$ (in the AC case) or periodic with period $T$ (in the AP case).


\section{Summary of results}

Our goal is to prove the following theorem.

\begin{thm} Let $(M,g)$ be a complete Riemannian $n$-manifold with a finite number of ends which are either AC or AP, and with scalar curvature satisfying $R \geq R_0 > 0$. Suppose that $\tau$ is a scalar function such that $\nabla \tau \in C^{0,\mu} _{-\delta} (TM)$ for some $0 < \mu < 1$, and $\sigma \in C^{1,\mu}$ is a symmetric 2-tensor with $||\sigma||_0$ sufficiently small relative to $R_0$ and $||\nabla \tau||^{1-n} _{0, -\delta}$. If $\sigma \not\equiv 0$, $\beta \geq \beta_0 > 0$, and $g$ admits no globally defined $L^2$ conformal Killing fields, then the system (\ref{eq:lcby}) has a solution.

\end{thm}

We will prove Theorem 2.1 by using a fixed point argument modeled on the ones by Holst, Nagy, and Tsogtgerel in \cite{HNT} and by Maxwell in \cite{M}. In those papers the authors construct global barriers to find solutions of (\ref{eq:lcby}) (with \cite{HNT} considering the non-vacuum case) for closed manifolds. Section 3 is dedicated to constructing suitable global barriers in the AC/AP case, and in section 4 we describe the corresponding fixed point construction.

From the discussion in the previous section, Theorem 2.1 implies that given any AC/AP metric $g$ with positive scalar curvature bounded away from zero and any symmetric 2-tensor $K$ whose trace $\tau$ satisfies the conditions of the theorem, we may find a solution $(\overline{g},\overline{K})$ to the constraint equations (\ref{eq:constr}) such that $\overline{g}$ is conformally equivalent to $g$. The metric $\overline{g}$ itself will not, in general, be AC/AP, leading us to ask whether this existence result can be extended to hold if we only assume that $g$ is \textit{conformally} AC/AP. In section 5 we analyze the operator $\Delta_\LL$ associated to a conformally AC metric and show that the constructions of sections 3 and 4 extend in a natural way to handle the conformally AC case. 

\begin{thm}
Let $(M,g)$ be a complete Riemannian $n$-manifold with a finite number of conformally AC ends. If $(M,g)$ is Yamabe positive and $\tau$, $\sigma$ are as in Theorem 2.1 with $||\sigma||_0$ sufficiently small relative to $||\nabla \tau||^{1-n} _{0, -\delta}$, then (\ref{eq:lcby}) has a solution.
\end{thm}

\noindent We extend this result to metrics which are conformally AP, but must then impose some additional restrictions on the metric. 

We now define the weighted Sobolev and H\"{o}lder norms used throughout this paper. For any $1 \leq p < \infty$, define the $W^{k,p} _\delta$-norm by
\begin{equation}
||X||_{W^{k,p} _\delta} = \bigg ( \sum\limits_{j \leq s} \int_M |\nabla^j X|^p e^{-p \delta t} dV_g \bigg )^{1/p}.
\end{equation}

\noindent The smooth function $t$ is said to be ``radial'' in the sense that on each end there are positive constants $C_1 < C_2$ such that
\begin{equation*}
C_1 t \leq 1 + \mathrm{dist} (\cdot, \partial E_\ell) \leq C_2 t.
\end{equation*}

\noindent On each end $E_\ell$, the set $\{t = a\}$ is diffeomorphic to $N_\ell$ for any $a > 0$, and we define $t$ to be identically zero on the compact part of the manifold which is disjoint from the ends. In keeping with the convention of \cite{CMP}, we will denote by $H^k _\delta$ the function space $W^{k,2} _\delta$. For any $q \in M$, define $B_1 (q)$ to be the ball of radius 1 about $q$ (where we assume without loss of generality that the radius of injectivity for $M$ is greater than 1). We define the H\"{o}lder norms as
\begin{equation}
||X||_{k,\mu; B_1 (q)} = \sum\limits^k _{i=0} \sup\limits_{B_1 (q)} |\nabla^i X| + \sup\limits_{x,y \in B_1 (q)} \dfrac{|\nabla^k X (x) - \nabla^k X(y)|}{d_g (x,y,)^\mu}
\end{equation}

\begin{equation}
||X||_{k,\mu} = \sup\limits_{q \in M} ||X||_{k,\mu; B_1 (q)},
\end{equation}

\noindent and also the weighted H\"{o}lder norms by
\begin{equation}
||X||_{k,\mu,\delta} = ||e^{-\delta t} X||_{k,\mu}.
\end{equation}

The operator $\Delta_\LL$ which appears in (\ref{eq:lcby}) is called the \textit{conformal vector Laplacian}. This elliptic operator and its properties on AC/AP manifolds were studied extensively in \cite{CMP}, and the results in that paper are central to the proof of our main theorem. In particular, in order to carry out the fixed point argument analogous to that in \cite{M}, we will need some information about the solvability of the equation $\Delta_\LL W = V$ when the underlying metric is AC/AP. Let us first define $\mathscr{Y}_\ell$ to be the set of all globally bounded vector fields which restrict to conformal Killing fields on the AC/AP end $E_\ell$. In the AP case, assume every $Y \in \mathscr{Y}_\ell$ also satisfies $\Gamma_* Y = \mathrm{Re}$ $e^{i\theta} Y$ for some $\theta \in \RR$, where $\Gamma_*$ is the pull-back of a generator $\Gamma$ of the group of deck transformations on the exactly periodic end (see \S 5 in \cite{CMP}). For each $\ell$ we define a smooth cutoff function $\chi_\ell$ which is identically one on $E_\ell$ and vanishes on every other end. We let $\mathscr{Y} = \oplus \{ \chi_\ell Y : Y \in \mathscr{Y}_\ell\}$. Theorem 6.1 in \cite{CMP} then spells out the Fredholm properties of $\Delta_\LL$ acting between weighted Sobolev spaces:

\begin{thm}
Let (M,g) be a Riemannian $n$-manifold with a finite number of ends which are either asymptotically cylindrical or asymptotically periodic. Suppose further that there is no nontrivial globally defined solution to $\Delta_\LL Y = 0$ which lies in $L^2(TM)$. Then there exists a number $\delta_* > 0$ such that if $0 < \delta < \delta_*$, then
\begin{equation}
\Delta_\LL : H^{k+2} _\delta (TM) \rightarrow H^k _\delta (TM)
\end{equation}

\noindent is surjective and the map
\begin{equation}
\Delta_\LL : H^{k+2} _{-\delta} (TM) \rightarrow H^k _{-\delta} (TM)
\end{equation}

\noindent is injective for every $k \geq 0$. Moreover, if $0 < \delta < \delta_*$, then for all $k \geq 0$ the map
\begin{equation}
\Delta_\LL : H^{k+2} _{-\delta} (TM) \oplus \mathscr{Y} \rightarrow H^k _{-\delta} (TM)
\end{equation}

\noindent is surjective with finite dimensional nullspace.

\end{thm}

\noindent It then follows from a standard parametrix construction that we can find a bounded generalized inverse $G : H^k _{-\delta} (TM) \rightarrow H^{k+2} _{-\delta} (TM) \oplus \mathscr{Y}$ satisfying $G \circ \Delta_\LL = I$, and analogous results can also be proved for weighted H\"{o}lder spaces (and weighted $L^p$-Sobolev spaces) using only a slight modification of the proof of Theorem 2.3. To be more specific, the authors of \cite{CMP} cite the results in \cite{Maz} (for the AC case) and \cite{MPU} (for the AP case) to construct parametrices for the operator $\Delta_\LL$ acting between weighted Sobolev spaces of functions on $M$. These same operators are also bounded maps between the appropriate H\"older spaces, so the arguments in \cite{Maz} and \cite{MPU} apply equally well in the H\"older case. 


\section{Global barriers}

The system (\ref{eq:lcby}) consists of a scalar equation and a vector equation. The former is commonly known as the \textit{Lichnerowicz equation} and the latter as the \textit{momentum constraint equation}. Given a positive function $\phi$, let $W_\phi$ be the solution to the corresponding momentum constraint equation (we henceforth drop the tildes from our notation). Now suppose $\phi_+$ is a positive function with the property that for any $\phi$ satisfying $0 < \phi \leq \phi_+$, we have $\Lich_\phi (\phi_+) \leq 0$, where we are using the notation
\begin{equation}
\Lich_\phi (\theta) := \Delta \theta - c_n R \theta - \beta \theta ^\alpha + c_n |\sigma + \DD W_\phi|^2 \theta ^{-\gamma}.
\end{equation}

\noindent Note that we have dropped the subscript of ``$g$'' from all metric-dependent operators and functions since there is no ambiguity going forward. A function $\phi_+$ with this property is called a \textit{global supersolution} of the system (\ref{eq:lcby}). If the inequality $\Lich_\phi (\phi_+) \leq 0$ only holds in the weak sense, we say that $\phi_+$ is a \textit{weak global supersolution}. Given a weak global supersolution, we define a continuous function $\phi_- > 0$ to be a \textit{weak global subsolution} if, for all $\phi_- \leq \phi \leq \phi_+$, we have $\Lich_\phi (\phi_-) \geq 0$ in the weak sense. 

The utility of global sub-/supersolutions is realized in the following proposition, a proof of which can be found in \cite{CM}.

\begin{prop}
Let $(M,g)$ be a smooth Riemannian manifold and $F$ a locally Lipschitz function. Suppose that $\underline{\phi} \leq \overline{\phi}$ are continuous functions which satisfy
\begin{equation*}
\Delta_g \underline{\phi} \geq F(x,\underline{\phi}), \hspace{.5cm} \Delta_g \overline{\phi} \leq F(x,\overline{\phi})
\end{equation*}

\noindent weakly. Then there exists a smooth function $\phi$ on $M$ such that
\begin{equation*}
\Delta_g \phi = F(x,\phi), \hspace{.5cm} \underline{\phi} \leq \phi \leq \overline{\phi}.
\end{equation*}
\end{prop}

\noindent Thus given a pair of global sub-/supersolutions and a function $\phi$ satisfying $\phi_- \leq \phi \leq \phi_+$, we can find a solution $\tilde{\phi}$ to the Lichnerowicz equation with $W = W_\phi$. If $\tilde{\phi} = \phi$, then this function solves the system (\ref{eq:lcby}). 

It would be quite useful to know whether the solution $\tilde{\phi}$ guaranteed by Proposition 3.1 is unique.  While we have no reason to suspect that uniqueness holds for an arbitrary pair of global sub-/supersolutions, one may show that uniqueness holds in certain special cases. For example, if we require that $\overline{\phi} - \underline{\phi} \leq Ce^{-\omega t}$ for some constant $C$ and look for solutions in an appropriate H\"{o}lder space, uniqueness follows from a slight adaptation of an argument given in \cite{CIP}, which contains the analogous uniqueness result when $M$ is asymptotically Euclidean. 

\begin{prop}
Let $\phi_1, \phi_2 \in C^{2,\mu}$ be a pair of positive solutions of the Lichnerowicz equation
\begin{equation} \label{eq:lich}
\Delta \phi - c_n R \phi - \beta \phi^\alpha + c_n |\tilde{\sigma}|^2 \phi^{-\gamma} = 0
\end{equation}
which are bounded away from zero, and suppose the underlying manifold $(M,g)$ and the function $\beta$ are as in Theorem 2.1 and $|\tilde{\sigma}|^2 \rightarrow \mr{\sigma}_\ell ^2$ on $E_\ell$ at the rate $e^{-\omega t}$ for some $\omega > 0$. If $\phi_2 - \phi_1 \in C^{2,\mu} _{-\omega}$, then $\phi_1 = \phi_2$.
\end{prop}

\noindent \textit{Proof sketch.} Defining new metrics on $M$ by $g_i = \phi_i ^{4/(n-2)} g$, so that $g_2 = (\phi_2 \phi_1 ^{-1} )^{4/(n-2)} g_1$, an argument nearly identical to the proof of Theorem 8.3 in \cite{CIP} proves that we have
\begin{equation}
(\Delta_{g_1} - \lambda)(\phi_2 \phi_1 ^{-1} - 1) = 0
\end{equation}

\noindent for some nonnegative continuous function $\lambda$. One easily checks that, under the given hypothesis, $\phi_2 \phi_1 ^{-1} - 1 \in C^{2,\mu} _{-\omega} (M)$. Since $\Delta_{g_1} - \lambda$ is seen to be injective as a map $C^{2,\mu} _{-\omega}(M) \rightarrow C^{0,\mu} _{-\omega}(M)$ by the maximum principle, we conclude that $\phi_2 \phi_1 ^{-1} - 1 = 0$, or $\phi_1 = \phi_2$ as claimed.\qed

\vspace{.5cm}

We note that Chru\'{s}ciel and Mazzeo proved in \cite{CM} that one may always solve (\ref{eq:lich}) with the given hypotheses by constructing sub-/supersolutions $\underline{\phi} \leq \overline{\phi}$ which depend on $\tilde{\sigma}$ and both approach a limit function $\mr{\phi}$ on the ends at a rate of $e^{-\omega t}$. Proposition 3.2 implies that the solution they obtain is unique among functions with the same asymptotic limit. If we thus fix limiting functions $\mr{\phi}, \mr{\beta}$ and $\mr{\sigma}^2 _\ell$, and let $\Sigma$ be the set of tensors $\tilde{\sigma}$ satisfying the hypothesis of Proposition 3.2, then Theorem 4.7 in \cite{CM} gives us a well-defined map $\mathcal{Q}: \Sigma \rightarrow C^{2,\mu}$ sending $\tilde{\sigma}$ to the unique solution of (\ref{eq:lich}) with asymptotic limit $\mr{\phi}$ (note that the proof of this theorem implies the hypothesis ``$\mr{\tilde{\sigma}}^2 > 0$'' can be weakened to ``$\mr{\tilde{\sigma}}^2 \not\equiv 0$''). In particular, if we construct a pair of \textit{global} sub-/supersolutions $\phi_- \leq \phi_+$ such that $\phi_{\pm} \rightarrow \mr{\phi}$ on the ends at the rate $e^{-\omega t}$, then for any $\phi_- \leq \phi \leq \phi_+$, we have $\phi_- \leq \mathcal{Q}(\sigma + \DD W_\phi) \leq \phi_+$ for some fixed $\sigma$ defined as in Theorem 2.1. This observation will be crucial to our fixed point construction below. To find such global barriers, we first find a constant global supersolution by applying what is essentially the argument in Proposition 14 of \cite{M}, which is itself a variant of an argument given in \cite{HNT}. However, as we are working in the noncompact case, we first establish some control over the behavior of $\DD W_\phi$ on the ends in terms of $||\phi||_0$.

\begin{lem}
Let $\phi$ be a bounded continuous function on $M$ and $\nabla \tau \in C^0 _{-\delta'} (TM)$ for some positive $\delta' < \delta_*$. If $W_\phi$ is the solution of the corresponding momentum constraint equation, then for any $\delta < \delta'$, there exists some constant $K$ which does not depend on $\phi$ or $\tau$ such that the following pointwise estimate holds:
\begin{equation}
|\DD W_\phi | \leq K ||\nabla \tau||_{0,-\delta'} ||\phi||_0 ^{\gamma-1} e^{-\delta t}.
\end{equation}
\end{lem}

\proof First note that
\begin{equation} \label{eq:ineq1}
|\DD W_\phi | = e^{-\delta t} |e^{\delta t} \DD W_\phi | \leq e^{-\delta t} ||\DD W_\phi ||_{0,-\delta}.
\end{equation}

\noindent Since $\phi^{\gamma-1} \nabla \tau \in C^0 _{-\delta'}$, for any $\delta < \delta'$ and $p \geq 1$, $\phi^{\gamma-1} \nabla \tau \in L^p _{-\delta} (TM)$. Applying the $L^p$-Sobolev version of Theorem 2.3, we see that the solution to the corresponding momentum constraint equation $W_\phi$ belongs to $W^{2,p} _{-\delta} \oplus \mathscr{Y}$ and that, endowing the finite-dimensional subspace $\mathscr{Y}$ with the $W^{2,p} _\delta$ norm (which is finite by the global boundedness of elements in $\mathscr{Y}$ and thus equivalent to any other choice of norm), the existence of a generalized inverse for $\Delta_\LL$ implies
\begin{equation}
||W_\phi||_{W^{2,p} _{-\delta} \oplus \mathscr{Y}} \leq C ||\phi^{\gamma-1} \nabla \tau || _{L^p _{-\delta}} \leq C ||\phi||^{\gamma-1} _0 ||\nabla \tau||_{L^p _{-\delta}}.
\end{equation}

\noindent Choosing $p$ sufficiently large and applying standard embedding arguments, the inequalities above imply
\begin{equation} \label{eq:ineq2}
||W_\phi||_{C^{1,\mu} _{-\delta} \oplus \mathscr{Y}} \leq C' ||\phi||^{\gamma-1} _0 ||\nabla \tau ||_{L^p _{-\delta}}.
\end{equation}

\noindent Moreover, it follows easily from the definition of $L^p _{-\delta}$-norm that there is a constant $C_p$ such that $||\nabla \tau||_{L^p _{-\delta}} \leq C_p ||\nabla \tau||_{0,-\delta'}$. Finally, note that $\DD$ is a bounded map $C^{1,\mu} _{-\delta} \oplus \mathscr{Y} \rightarrow C^{0,\mu} _{-\delta}$, and we obviously have $||\DD W_\phi ||_{0, -\delta} \leq ||\DD W_\phi||_{C^{0,\mu} _{-\delta}}$. These observations along with (\ref{eq:ineq1}) and (\ref{eq:ineq2}) prove the assertion. \qed

\vspace{.5cm}

Using Lemma 3.3, finding a constant global supersolution becomes a relatively simple matter under the hypothesis of Theorem 2.1.

\begin{prop}
Suppose, as in the hypothesis of Theorem 2.1, that there are constants $R_0, \beta_0$ which are positive lower bounds for the scalar curvature $R$ and $\beta$ respectively. Then if $||\sigma||_0$ is sufficiently small relative to $R_0$ and $||\nabla \tau||_{0, -\delta'} ^{1-n}$, the system (\ref{eq:lcby}) admits a constant global supersolution.
\end{prop}

\proof Let $K_\tau = K ||\nabla \tau ||_{0,-\delta'}$ where $K$ is as in Lemma 3.3, and let $\epsilon_+$ be some positive constant. We will argue that if $\epsilon_+ > 0$ is sufficiently small, then it is a global supersolution of (\ref{eq:lcby}). We have $R \geq R_0$, so for any $0 < \phi \leq \epsilon_+$ we have
\begin{align*}
\mathrm{Lich}_\phi(\epsilon_+) & = -c_n R \epsilon_+ - \beta \epsilon_+ ^\alpha + c_n |\mathcal{D}W_\phi + \sigma|^2 \epsilon_+ ^{-\gamma}\\
&\leq -c_n R \epsilon_+ + 2c_n |\mathcal{D}W_\phi|^2 \epsilon_+ ^{-\gamma} + 2c_n|\sigma|^2 \epsilon_+ ^{-\gamma} \\
&\leq -c_n R_0 \epsilon_+ + 2c_n ||\mathcal{D}W_\phi||_0 ^2 \epsilon_+ ^{-\gamma} + 2c_n ||\sigma||_0 ^2 \epsilon_+^{-\gamma}\\
&\leq -c_n R_0 \epsilon_+ + 2c_n K_\tau ^2 \epsilon_+ ^{2\gamma - 2 - \gamma} + 2c_n ||\sigma||_0 ^2 \epsilon_+^{-\gamma} \\
&= -c_n R_0 \epsilon_+ + 2c_n K_\tau ^2  \epsilon_+ ^{\alpha} + 2c_n ||\sigma||_0 ^2 \epsilon_+^{-\gamma}
\end{align*}

\noindent where we have used Lemma 3.3 and the fact that $\phi \leq \epsilon_+$ in the third ineqality, and the identity $\gamma - 2 = \alpha$ in the final equality. Examining the first two terms in the last line above, we see that if we choose 
\begin{equation*}
\epsilon_+ ^{\alpha-1} < \dfrac{R_0}{2K_\tau ^2},
\end{equation*}

\noindent then the total contribution of the first two terms in this last expression is negative. In particular, if we choose $\epsilon_+$ so small that
\begin{equation*}
\epsilon_+ ^{\alpha - 1} < \dfrac{R_0}{4 K_\tau ^2},
\end{equation*}

\noindent we find that the contribution of those terms is less than $-2c_nK^2 _\tau \epsilon_+ ^\alpha$. Hence if we require that 
\begin{align*}
||\sigma||_0 ^2 < K_\tau ^2 \epsilon_+ ^{\alpha + \gamma} \Rightarrow ||\sigma||_0 &< K_\tau \epsilon_+ ^{(\alpha + \gamma)/2} \\
&= K_\tau (\epsilon_+ ^{\alpha - 1})^{(\alpha+1)/(\alpha-1)} \\
&< K_\tau (R_0/4K_\tau ^2)^{n/2} \\
&= 2^{-n} K_\tau ^{1-n} R_0 ^{n/2},
\end{align*}

\noindent the positive contribution of the final term above is so small that $\Lich_\phi (\epsilon_+) \leq 0$. The constant $\epsilon_+$ is thus a global supersolution on $M$. \qed 

\vspace{.5cm}

\noindent A constant global supersolution is not particularly desirable, for unless we construct a global subsolution which approaches this constant asymptotically, we would be able to say little about the asymptotics of a solution provided by Proposition 3.1. Because we would like to obtain better information on the asymptotics of solutions to (\ref{eq:lcby}), we would like to find a pair of global sub-/supersolutions which converge to the same limits on the ends, trapping the asymptotic behavior of the solution obtained by applying Proposition 3.1. To accomplish this, we use the constant global supersolution constructed above to define a limiting function $\mr{\phi}_\ell$ on each $N_\ell$, and then construct an appropriate pair of global sub-/supersolutions to guarantee that the solution of the Lichnerowicz equation between these approaches $\mr{\phi}_\ell$ asymptotically on $E_\ell$. With this goal in mind, consider the following construction. As in \cite{CM}, we define the \textit{reduced Lichnerowicz equation} to be the semilinear equation on each end given by
\begin{equation} \label{eq:redlich}
\Delta_{\mr{g}_\ell} \mr{\phi}_\ell - c_n \mr{R}_\ell \mr{\phi}_\ell - \mr{\beta}_\ell \mr{\phi}_\ell ^\alpha + c_n \mr{\sigma}_\ell ^2 \mr{\phi}_\ell ^{-\gamma} = 0.
\end{equation}

\noindent Here $\mr{g}_\ell$ is either the metric induced by $g$ on the ``cross-sectional'' manifold $N_\ell$ in the limit as $t \rightarrow \infty$ in the AC case, or the metric on $S^1 \times N_\ell$ given by the restriction of $g$ in the limit to its period in the AP case. $\mr{R}_\ell$ is the scalar curvature associated to this metric, and $\mr{\beta}_\ell$ is the constant function which is the asymptotic limit of $\beta$. Note that in the AC case this is \textit{not} the same as the Lichnerowicz equation on $N_\ell$ since the dimensional constants in the two equations differ.

Due to our assumptions on $R$, $\sigma$, and $\beta$, one easily checks that $\epsilon_+$ is a supersolution to the reduced Lichnerowicz equation on either $N_\ell$ or $S^1 \times N_\ell$. Moreover, a bounded subsolution $tu_\ell$ of this equation is constructed in \cite{CM}, where $t$ is any sufficiently small positive number. We may thus choose $t < \epsilon_+ (\max_\ell \max_{N_\ell} |u_\ell|)^{-1}$ so that Proposition 3.1 provides a solution $\mr{\phi}_\ell \leq \epsilon_+$ of (\ref{eq:redlich}). We then replace $\epsilon_+$ with $\epsilon' _+ > \epsilon_+$ where $\epsilon'_+$ is so small that $\epsilon'_+$ is still a global supersolution. Continuing to call this new global supersolution ``$\epsilon_+$'', we thus have $\mr{\phi}_\ell < \epsilon_+$ for each $\ell$, and we extend its definition to all of $E_\ell$ by defining it to be translation invariant in the AC case and invariant under $t$-translations by multiples of $T$ in the AP case. We extend these functions using cutoffs to a positive $C^{2,\mu}$ function $\mr{\phi}$ on all of $M$.

Given $\delta_* > 0$ as in Theorem 2.3, assume that $\nabla \tau \in C^{0,\mu} _{-\delta'}$ where $0 < \delta' < \delta_*$. Now choose any positive $\delta < \delta'$, select some positive number $\nu < \delta/(2\gamma - 2)$ and let $u$ be the unique solution in $C^{2,\mu} _{-\nu}(M)$ of the equation
\begin{equation} \label{eq:decay1}
\Delta u - c_n R u = -e^{-\nu t}.
\end{equation}

\noindent We ultimately define a global supersolution $\phi_+$ to equal $\epsilon_+$ on some compact set $\KK$ and $\min\{ \epsilon_+, \mr{\phi} + bu\}$ outside of $\KK$ where $b$ is a large constant. This is a weak global supersolution provided that $\mr{\phi} + bu$ is a global supersolution on $M \setminus \KK$ and that $\mr{\phi} + bu \geq \epsilon_+$ on $\KK$. We will next show that, with the usual assumption that $\beta$ is bounded below away from zero, we may find a compact set $\KK \subset M$ such that $\mr{\phi} + bu$ is a global supersolution outside of $\KK$ for any sufficiently large $b$. If we then choose $b$ large enough so that $\mr{\phi} + bu > \epsilon_+$ on all of $\KK$, $\phi_+ = \min \{\epsilon_+, \mr{\phi} + bu\}$ is a continuous weak global supersolution of (\ref{eq:lcby}) on $M$ which asymptotically approaches $\mr{\phi}$ on the ends.

\begin{prop}
Let $0 < \nu < \delta/(2\gamma - 2)$ be so small that $\Lich_0 (\mr{\phi})$ decays faster than $e^{-\nu t}$. Let $u \in C^{2,\mu}_{-\nu}$ be defined as in (\ref{eq:decay1}) and suppose that $\beta \geq \beta_0 > 0$ where $\beta_0$ is a constant. Then there exists a compact set $\mathcal{K} \subset M$ such that $\mathring{\phi} + bu$ is a global supersolution on $M \setminus \mathcal{K}$ for any sufficiently large $b$.
\end{prop}

\proof First note that the maximum principle and the asymptotics of the metric imply that there are constants $k_1$ and $k_2$ such that $k_1 e^{-\nu t} \leq u \leq k_2 e^{-\nu t}$. If $0 < \phi \leq \mathring{\phi} + bu$, we have  
\begin{equation*}
\mathrm{Lich}_\phi (\mathring{\phi} + bu) = (\Delta - c_n R)\mr{\phi} - be^{-\nu t} - \beta (\mr{\phi} + bu)^\alpha + c_n |\DD W_\phi + \sigma|^2 (\mr{\phi} + bu)^{-\gamma}. 
\end{equation*}

\noindent Because $\alpha > 1$, by convexity we may bound this expression above by
\begin{equation*}
(\Delta - c_n R)\mr{\phi} - be^{-\nu t} - \beta \mr{\phi}^\alpha - \beta (bu)^\alpha + c_n (|\DD W_\phi| + |\sigma|)^2 (\mr{\phi} + bu)^{-\gamma}.
\end{equation*}

\noindent We next use the fact that $\gamma > 1$ to note that $(\mr{\phi} + bu)^{-\gamma} < \min\{\mr{\phi}^{-\gamma}, (bu)^{-\gamma}\}$ to bound the previous expression above by
\begin{multline}
(\Delta - c_n R)\mr{\phi} - \beta \mr{\phi}^\alpha + c_n |\sigma|^2 \mr{\phi}^{-\gamma} - be^{-\nu t} - \beta_0 (bu)^\alpha \\
 + c_n (|\DD W_\phi|^2 + 2 |\DD W_\phi| |\sigma|) (bu)^{-\gamma}.
\end{multline}

\noindent The first three terms in this expression equal $\Lich_0 (\mr{\phi})$, so we may use Lemma 3.3 to bound this last expression above by
\begin{align*}
&\quad \Lich_0 (\mathring{\phi}) - be^{-\nu t} - \beta_0 b^\alpha u^\alpha + c_n|\mathcal{D}W_\phi|^2 b^{-\gamma} u^{-\gamma} + 2c_n |\mathcal{D}W_\phi||\sigma| b^{-\gamma} u^{-\gamma} \\
&\leq \mathrm{Lich}_0 (\mathring{\phi}) - be^{-\nu t} - c_1 \beta_0 b^\alpha e^{-\alpha \nu t} + c_2 ||\mathring{\phi} + bu||_0 ^{2\gamma -2} b^{-\gamma} e^{(\nu\gamma - 2\delta)t} \\
& \quad + c_3 ||\mathring{\phi} + bu||_0 ^{\gamma - 1} ||\sigma||_0 b^{-\gamma} e^{(\nu\gamma - \delta)t}
\end{align*}

\noindent where $c_1 = k_1 ^\alpha$, $c_2 = c_n k_1 ^{-\gamma} K_\tau ^2$, and $c_3 = 2c_n k_1 ^{-\gamma} K_\tau$. Now choose $b$ so large that $||\mathring{\phi}/b + u ||_0 \leq 2||u||_0$. Then letting $\tilde{c}_2 = c_2 (2 ||u||_0) ^{2\gamma - 2}$ and $\tilde{c_3} = c_3 (2 ||u||_0)^{\gamma-1}$, we see that the final expression above is bounded above by
\begin{equation*}
\mathrm{Lich}_0 (\mathring{\phi}) - be^{-\nu t} - c_1 \beta_0 b^\alpha e^{-\alpha\nu t} + \tilde{c}_2 b^\alpha e^{(\gamma\nu - 2\delta)t} + \tilde{c}_3 ||\sigma||_0 b^{-1} e^{(\gamma\nu - \delta)t}.
\end{equation*}

\noindent where we have used that $\gamma - 2 = \alpha$. One easily sees that the final two terms in the expression above decay faster than the middle term since $\nu < \delta/(2\gamma - 2)$ by assumption. By hypothesis, there is some $t_0$ such that $\mathrm{Lich}_0 (\mathring{\phi}) < e^{-\nu t}$ whenever $t > t_0$, so we may choose some $t_1 > t_0$ such that
\begin{equation*}
-c_1 \beta_0 b^\alpha e^{-\alpha\nu t} + \tilde{c}_2 b^\alpha e^{(\nu\gamma - 2\delta)t} + \tilde{c}_3 ||\sigma||_0 b^\alpha e^{(\nu t - \delta)t} < 0.
\end{equation*}

\noindent whenever $t > t_1$. Clearly $\tilde{c}_3 b^{-1} ||\sigma||_0 e^{(\nu t - \delta)t} \leq \tilde{c}_3 b^\alpha ||\sigma||_0 e^{(\nu t - \delta)t}$, so we may set $\mathcal{K} = \{ t \leq t_1 \}$ to prove the proposition. \qed 

\vspace{.5cm}

We may similarly construct a weak global subsolution which is asymptotic to $\mr{\phi}$. Because for every $\phi \leq \phi_+$ we have that $|\mathcal{D}W_\phi| \leq K_\tau ||\phi_+||^{\gamma-1} _0 e^{-\delta t}$ by Lemma 3.3, outside of some large compact set of the form $\{ t \leq t_2 \}$ (which we will also denote by $\mathcal{K}$) we have
\begin{equation*}
\mathrm{Lich}_0 (\mathring{\phi}) + e^{-\nu t} - c_n |\mathcal{D}W_\phi|(|\mathcal{D}W_\phi | + 2 |\sigma|)(\min_M \mathring{\phi})^{-\gamma} \geq 0
\end{equation*}

\noindent where $\nu$ is as in the previous proposition. Now let $v$ be the unique solution to 
\begin{equation*}
\begin{cases} (\Delta - c_n R )v = -e^{-\nu t} \\ v|_{\partial \mathcal{K}} = \mathring{\phi}|_{\partial \mathcal{K}} \end{cases}
\end{equation*}

\noindent on $\overline{M \setminus \mathcal{K}}$ and define a function $\varphi_-$ on $M \setminus \mathcal{K}$ by $\varphi_- = \mathring{\phi} - v$ (note that $v$ is positive and hence $\varphi_- < \mathring{\phi}$). Wherever $\varphi_-$ is positive on $M \setminus \mathcal{K}$, we have
\begin{align*}
\Lich_\phi (\varphi_-) &= (\Delta - c_n R)\varphi_- - \beta \varphi_- ^\alpha + c_n|\mathcal{D}W_\phi + \sigma|^2 \varphi_- ^{-\gamma}\\
&\geq (\Delta - c_n R)\mr{\phi} + e^{-\nu t} - \beta \mr{\phi}^\alpha + c_n|\DD W_\phi + \sigma|^2 \mr{\phi}^{-\gamma}\\
&\geq (\Delta - c_n R)\mathring{\phi} - \beta \mathring{\phi} ^\alpha + c_n|\sigma|^2 \mathring{\phi} ^{-\gamma} + e^{-\nu t} \\
&\quad +c_n |\mathcal{D}W_\phi|(|\mathcal{D}W_\phi| - 2|\sigma|)\mathring{\phi}^{-\gamma} \\
&\geq \mathrm{Lich}_0 (\mathring{\phi}) + e^{-\nu t} - c_n |\mathcal{D}W_\phi|(|\mathcal{D}W_\phi | + 2 |\sigma|)(\min_M \mathring{\phi})^{-\gamma}\\
&\geq 0
\end{align*}

\noindent where the final inequality follows from the definition of $\mathcal{K}$. Since $v$ decays on the ends and $\mathring{\phi}$ is $t$-invariant or periodic on the ends, we know that $\varphi_- = \mathring{\phi} - v$ is strictly positive outside of some compact set containing $\mathcal{K}$. Define such a set $\mathcal{K}' \supset \mathcal{K}$ where $\varphi_- > \frac{1}{2} \min_M \mathring{\phi}$ on $M \setminus \mathcal{K}'$. We then let $\eta$ be the solution on $\mathcal{K}'$ to the following boundary value problem:
\begin{equation*}
\begin{cases} (\Delta - c_n R - \beta)\eta = 0 \\ \eta|_{\partial \mathcal{K}'} = \frac{1}{2} \min \{ 1, \min_M \mathring{\phi} \} \end{cases}.
\end{equation*}

\noindent One easily checks that $\eta$ is a global subsolution for the Lichnerowicz equation on $\mathcal{K}'$, is positive by the strong maximum principle, and is less than $\varphi_-$ on $\partial \mathcal{K}'$. We thus conclude that $\phi_- = \max \{ \varphi_-, \eta\}$ is a weak global subsolution for the Lichnerowicz equation on $M$.


\section{Fixed point argument}

The asymptotics of our global sub-/supersolution pair $\phi_- \leq \phi_+$ suggest that we look for a solution of the coupled system (\ref{eq:lcby}) with the form $\mr{\phi} + \psi$ where $\psi \in C^{2,\mu} _{-\nu}(M)$. Namely, if we treat $\mr{\phi}$ as fixed, we can write (\ref{eq:lcby}) as a nonlinear elliptic system for $(\psi,W)$. Similar to the approach in \cite{HNT} and \cite{M}, we will find a solution for this system by applying Schauder's fixed point theorem to a well-chosen function on a weighted H\"{o}lder space. Let us first state the fixed point theorem we have in mind.

\begin{thm}
Let $X$ be a Banach space, and let $U \subset X$ be a non-empty, convex, closed, bounded subset. If $T: U \rightarrow U$ is a compact operator, then there exists a fixed point $u \in U$ such that $T(u) = u$.
\end{thm}

A proof of Theorem 4.1 can be found in \cite{I}. Let $0 < \nu' < \nu$. We will find a solution $(\tilde{\psi},W_{\tilde{\psi}})$ to the $(\psi,W)$-system as a fixed point of a composition of suitably chosen solution operators on the set
\begin{equation}
U = \{ \psi \in C^0 _{-\nu'} : \phi_- - \mr{\phi} \leq \psi \leq \phi_+ - \mr{\phi} \}.
\end{equation}

\noindent This set clearly meets all the criteria of Theorem 4.1. Below we define a map $T$ and show that it preserves $U$.

Let us first define a map $\mathcal{W} : U \rightarrow C^{1,\mu} _{-\delta} (TM) \oplus \mathscr{Y}$ which sends $\psi$ to the vector field
\begin{equation}
W_\psi := G(n^{-1}(n+1)(\mr{\phi} + \psi)^{\gamma-1} \nabla \tau),
\end{equation}

\noindent where $G$ is a bounded generalized inverse for the conformal vector Laplacian $\Delta_\LL$ (the existence of $G$ follows from the H\"{o}lder space version of Theorem 2.3). We next define a map $\mathcal{S}_\sigma : C^{0,\mu} _{-\delta} (S^2 _0 (M)) \rightarrow C^{2,\mu} _{-\nu}(M)$ by 
\begin{equation*}
\pi \mapsto \mathcal{Q}(\sigma + \pi) - \mr{\phi}
\end{equation*}

\noindent where $\mathcal{Q}(\tilde{\sigma})$ is defined to be the unique solution $\phi_- \leq \tilde{\phi} \leq \phi_+$ of
\begin{equation}
\Delta \phi - c_n R \phi - \beta \phi^\alpha + c_n |\tilde{\sigma}|^2 \phi^{-\gamma} = 0
\end{equation}

\noindent which satisfies $\mathcal{Q}(\tilde{\sigma}) \rightarrow \mr{\phi}$ on the ends. $\mathcal{Q}$ is well-defined by the discussion in section 3. We show that $\mathcal{S}_\sigma$ is continuous using an argument very similar to the proof of Proposition 13 in \cite{M}.

\begin{lem}
The map $\mathcal{S}_\sigma : C^{0,\mu} _{-\delta} (S^2 _0 (M)) \rightarrow C^{2,\mu} _{-\nu}(M)$ is continuous.
\end{lem}

\noindent Note that here the domain of $\mathcal{S}_\sigma$ is the set of symmetric 2-tensors in the indicated weighted H\"{o}lder space.

\proof We invoke the implicit function theorem (see \cite{S}, for example). We define a map $F : C^{2,\mu} _{-\nu}(M) \times C^{0,\mu} _{-\delta}(S^2 _0 (M)) \rightarrow C^{0,\mu} _{-\nu}(M)$ by 
\begin{equation*}
F(\psi,\pi) = (\Delta - c_n R)(\mr{\phi} + \psi) - \beta (\mr{\phi} + \psi)^\alpha + c_n |\sigma + \pi|^2 (\mr{\phi} + \psi)^{-\gamma},
\end{equation*}

\noindent so that $F(\mathcal{S}_\sigma (\pi), \pi) = 0$ by definition. The Fr\'{e}chet derivative $DF_{(\psi,\pi)} (h,k)$ of $F$ at the point $(\psi, \pi)$ acting on pairs $(h,k) \in C^{2,\mu} _{-\nu}(M) \times C^{0,\mu} _{-\delta}(S^2 _0 (M))$ is given by 
\begin{equation*}
(\Delta - c_n R)h - \alpha \beta (\mr{\phi} + \psi)^{\alpha-1} h + 2c_n \langle \sigma + \pi, k \rangle (\mr{\phi} + \psi)^{-\gamma} - c_n \gamma |\sigma + \pi|^2 (\mr{\phi} + \psi)^{-\gamma - 1} h.
\end{equation*}

\noindent In particular,
\begin{equation}
DF_{(\psi,\pi)} (h,0) = \big [ \Delta - (c_n R + \alpha \beta (\mr{\phi} + \psi)^{\alpha-1} + c_n \gamma |\sigma + \pi|^2 (\mr{\phi} + \psi)^{-\gamma -1}) \big ] h.
\end{equation}

\noindent Hence given any pair $(\psi,\pi)$ for which $\mr{\phi} + \psi > 0$ everywhere, which is certainly true for all $\psi$ in the image of $\mathcal{S}_\sigma$, we see that $DF_{(\psi,\pi)} : C^{2,\mu} _{-\nu} (M) \rightarrow C^{0,\mu} _{-\nu} (M) $ is an isomorphism. Since it is obviously continuous in $(\psi,\pi)$, the implicit function theorem implies $\mathcal{S}_\sigma$ is continuous in an open neighborhood of $\pi$. \qed

\vspace{.5cm}

Since Proposition 3.1 guarantees that the range of $\mathcal{S}_\sigma \circ \DD \circ \mathcal{W}$ lies in the set $\{ \psi \in C^{2,\mu} _{-\nu} : \phi_- - \mr{\phi} \leq \psi \leq \phi_+ - \mr{\phi} \}$, we are now in a position to prove Theorem 2.1.

\vspace{.5cm}

\noindent \textit{Proof of Theorem 2.1.} With $U$ defined as above, define $T$ to be the composition of $\mathcal{S} \circ \DD \circ \mathcal{W}$ with the natural compact embedding $C^{2,\mu} _{-\nu}(M) \hookrightarrow C^0 _{-\nu'}(M)$. $T$ and $U$ thus satisfy the hypotheses of Theorem 4.1, and so $T$ has a fixed point $\tilde{\psi}$. By construction, $(\mr{\phi} + \tilde{\psi}, W_{\mr{\phi} + \tilde{\psi}})$ is a solution to the coupled system (\ref{eq:lcby}). \qed


\section{The Conformally AC Case}

Thus far we have assumed that our Riemannian metric is only asymptotically cylindrical or asymptotically periodic. We have deferred our discussion of the problem of solving (\ref{eq:lcby}) for the conformally AC/AP case because, at least in the conformally AC case, the solution requires only a slight modification of the arguments above. We shall focus on this case first. Namely, choose some $\tilde{\phi}$ which satisfies $\tilde{\phi} - \mr{\phi} \in C^{2,\mu} _{-\nu}$ and let $g$ be an AC metric on $M$. We then define a conformally AC metric $\tilde{g} = \tilde{\phi}^{4/(n-2)} g$ and seek to solve (\ref{eq:lcby}) with respect to this metric.

\subsection{Indicial roots of $\Delta_\LL$ for conformally AC metrics}

First note that if $g$ were exactly cylindrical and $\tilde{\phi} = \mr{\phi}$, a straightforward computation reveals that the associated conformal vector Laplacian $\tilde{\Delta}_\LL$ can be expressed with respect to the conformal Killing operator of the exactly cylindrical metric via
\begin{equation} \label{eq:cvltrans}
(\tilde{\Delta}_\LL X)^i = \mr{\phi}^{-4/(n-2)} (\Delta_\LL X)^i - \dfrac{n}{2} \nabla_j (\mr{\phi}^{-4/(n-2)}) (\DD X)^{ij}.
\end{equation} 

\noindent As in this equation, throughout this section only the operators denoted with a tilde are defined with respect to the conformally AC metric $\tilde{g}$ while all others are defined with respect to the AC metric $g$. As $\mr{\phi}$ is bounded above and below, this operator is uniformly elliptic. Now return to the general case. Since $\tilde{g}$ decays exponentially to $\mr{\phi}^{4/(n-2)}g$, $\tilde{\Delta}_\LL := \tilde{\DD}^* \tilde{\DD}$ can be expressed as the operator in (\ref{eq:cvltrans}) (where all the operators on the right are defined with respect to the \textit{exactly} cylindrical metric) plus a perturbation term whose coefficients decay like $e^{-\omega' t}$ for some $\omega' > 0$. In particular, we may regard $\tilde{\Delta}_\LL$ as an elliptic $b$-operator so that, with the obvious change of coordinates $x = e^{-t}$, we may cite results from \cite{Maz} to analyze this operator just as in the proof of Theorem 2.3. 

To be specific, as in \cite{CMP} for each $N_\ell$ we may define the \textit{indicial family} $I_\lambda (\hat{\Delta}_\LL)$ to be the family of operators on the cylinder $\RR \times N_\ell$ given by $I_\lambda (\hat{\Delta}_\LL) = \mathcal{F} \circ \hat{\Delta}_\LL \circ \mathcal{F}^{-1}$, where $\mathcal{F}$ is the Fourier transform and $\hat{\Delta}_\LL$ is the operator on the right hand side of (\ref{eq:cvltrans}) with an exactly cylindrical underlying metric $g$. One easily checks that $I_\lambda (\hat{\Delta}_\LL)$ is a second-order elliptic operator depending polynomially on $\lambda$, and note that our definition of $I_\lambda (\hat{\Delta}_\LL)$ is equivalent to $I_{i \lambda} (\hat{\Delta}_\LL)$ in \cite{Maz}. The analytic Fredholm theorem implies that this operator is invertible away from a discrete set of complex numbers $\Lambda(\hat{\Delta}_\LL) = \{\lambda_j\}$ which we define to be the \textit{indicial roots} of the conformally AC operator $\tilde{\Delta}_\LL$. It follows from the results of \cite{Maz} that the indicial roots of an elliptic $b$-operator such as $\hat{\Delta}_\LL$ are precisely the decay rates of elements in its nullspace (but note that due to our choice of definition for $I_\lambda (\hat{\Delta}_\LL)$, only real $\lambda$ correspond to bounded solutions of $\hat{\Delta}_\LL Y = 0$), and there are only finitely many in any horizontal strip $\{a < \mathrm{Im} z < b \}$. This implies that we may choose $\delta_*$ so small that the only indicial roots in the strip $\{-\delta_* < \mathrm{Im} z < \delta_* \}$ are real. It then follows from Theorem 4.26 in \cite{Maz} and an argument analogous to the proof of Theorem 2.3 that if $\delta$ belongs to the punctured interval $(-\delta_*, \delta_*)\setminus\{0\}$, the map
\begin{equation}
\tilde{\Delta}_\LL : C^{k+2, \mu} _{\delta}(TM) \rightarrow C^{k,\mu} _{\delta}(TM)
\end{equation}

\noindent is Fredholm for any $k \geq 0$. Injectivity and surjectivity arguments can then be made exactly as in the proof of Theorem 2.3, and Theorem 7.14 in \cite{Maz} implies that the generalized inverse $G$ maps $C^{k, \mu} _{-\delta} (TM)$ into $C^{k+2, \mu} _{-\delta} (TM) \oplus \mathscr{Y}_1 \oplus \cdots \oplus \mathscr{Y}_m$ where $\mathscr{Y}_i$ is a subspace of vector fields whose restrictions to each end are periodic solutions of $\hat{\Delta}_\LL Y = 0$. However, we claim that these vector fields actually restrict to conformal Killing fields of the exactly conformally cylindrical metric on each end. For if $T_{i \ell}$ is the period of $Y \in \mathscr{Y}_i$ restricted to $E_\ell$, quotienting the cylinder by $T_{i \ell} \ZZ$ gives us a solution to $\hat{\Delta}_\LL Y = 0$ on the closed manifold $S^1 \times N_\ell$. We may thus integrate by parts to find that, in fact, $\hat{\DD} Y \equiv 0$, implying that $Y$ is a conformal Killing field on each end. The difference $\tilde{\DD} - \hat{\DD}$ is a first order operator with coefficients decaying like $e^{-\omega t}$, and this means that if we take $\delta < \omega$, the image of $\mathscr{Y}_1 \oplus \cdots \oplus \mathscr{Y}_m$ under $\tilde{\DD}$ belongs to $C^{k+1, \mu} _{-\delta} (S^2 _0 (TM))$. Hence the composition $\tilde{\DD} \circ G$ is a bounded map from $C^{k, \mu} _{-\delta} (TM)$ into $C^{k+1, \mu} _{-\delta} (S^2 _0 (TM))$. This mapping property implies that Lemma 3.3 holds in the conformally AC case, and so we may bound the magnitude of $\tilde{\DD} W_\phi$ in terms of $||\phi||_0$, allowing us to construct a pair of global sub-/supersolutions.

\subsection{Global barriers in the conformally AC case}

To prove Theorem 2.2, we must show that we can replace the condition $R \geq R_0 > 0$ with the weaker condition that $(M,g)$ is \textit{Yamabe positive}. This is to say that the Yamabe invariant of the conformal class $[g]$ is positive, recalling that the Yamabe invariant for a manifold with cylindrical ends is defined as
\begin{equation}
Y(M,[g]) = \inf\limits_{\substack{u \in C^\infty _0 \\ 0 \leq u \not\equiv 0}} \dfrac{\frac{1}{2} \int_M (|\nabla u |^2 + c_n R u^2)}{(\int_M u^{\frac{2n}{n-2}} )^{\frac{n-2}{n}}}.
\end{equation}

\noindent We have the following proposition, which is direct consequence of the proof of Proposition 4.6 in \cite{CM}:

\begin{prop}
Suppose that $(M,g)$ has conformally AC or AP ends, and that $(M,g)$ is Yamabe positive. Then there exists a positive function $u \in C^\infty (M)$ such that on each end $E_\ell$, $u \rightarrow u_\ell$ as $t \rightarrow \infty$, where $u_\ell \in C^\infty (N_\ell)$ in the AC case and $u_\ell \in C^\infty (S^1 \times N_\ell)$ in the AP case, such that $\check{g} = u^{4/(n-2)} g$ has $\check{R} \geq \check{R}_0 > 0$ everywhere.
\end{prop}

\noindent In particular, so long as $(M,g)$ is Yamabe positive, we may conformally transform $g$ to a metric which has scalar curvature bounded below by a positive constant. This fact allows us to construct global sub-/supersolutions much in the same way as in section 3.

\begin{prop}
Let $(M,g)$ be a manifold with conformally AC ends with $Y(M,[g]) > 0$, and suppose $\tau$ is as in Theorem 2.2. If $\vartheta$ is the conformal factor given by Propsition 5.1, then we may find some constant $\epsilon > 0$ such that $\epsilon \vartheta$ is a global supersolution of the system (\ref{eq:lcby}) if $||\sigma||_0$ is sufficiently small.
\end{prop}

\proof The proof is again essentially the same as in Proposition 14 of \cite{M}. From the identity $\Delta \vartheta - c_n R \vartheta = -c_n \check{R} \vartheta^\alpha$ we see that given any $\phi \leq \epsilon \vartheta$,
\begin{align*}
\Lich_\phi (\epsilon \vartheta) &= -\epsilon c_n \check{R} \vartheta^\alpha - \beta \epsilon^\alpha \vartheta^\alpha + c_n |\sigma + \DD W|^2 \epsilon^{-\gamma} \vartheta^{-\gamma} \\
&\leq -\epsilon c_n \check{R}_0 \vartheta^\alpha - \beta \epsilon^\alpha \vartheta^\alpha + 2c_n |\sigma|^2 \epsilon^{-\gamma} \vartheta^{-\gamma} + 2c_n |\DD W |^2 \epsilon^{-\gamma} \vartheta^{-\gamma}.
\end{align*}

\noindent Using Lemma 3.3 (which we showed to be valid in the conformally AC case), it is then easy to see that this quantity is bounded above by
\begin{equation*}
(-c_n \check{R}_0 \epsilon - \beta_0 \epsilon^\alpha)(\inf \vartheta)^\alpha + (2c_n ||\sigma||_0 ^2 \epsilon^{-\gamma} + 2c_n K_\tau ^2 \epsilon^\alpha  (\sup \vartheta)^{2\gamma - 2}) (\inf \vartheta)^{-\gamma}.
\end{equation*}

\noindent The rest of the argument proceeds exactly as in the proof of Proposition 3.4, though now the smallness condition on $||\sigma||_0$ will depend on $K_\tau, \check{R}_0, \inf \vartheta,$ and $\sup \vartheta$. \qed

\vspace{.5cm}

\noindent We again observe that the global supersolution constructed above asymptotically approaches a supersolution of the reduced Lichnerowicz equation on each end, and we may thus invoke the subsolution constructed in \cite{CM} and Proposition 3.1 to find a solution $\mr{\phi}_\ell$ on each end. Extend this solution to all of $M$ as above. In order to construct a global supersolution which is asymptotic to $\mr{\phi}$, we need a result analogous to Proposition 3.5.

\begin{prop}
Let $(M,g)$ and $\vartheta$ be as in the previous proposition. If $\nu > 0$ is sufficiently small, then we may find a compact set $\mathcal{K} \subset M$ such that $\mr{\phi} + b \vartheta e^{-\nu t}$ is a global supersolution on $M \setminus \mathcal{K}$ for any sufficiently large $b$.
\end{prop}

\proof One easily checks that there is a constant $A = A( n,g,||\vartheta||_{C^1})$ satisfying
\begin{equation*}
|2\nabla \vartheta \cdot \nabla e^{- \nu t} + \vartheta \Delta e^{-\nu t}| \leq \nu A e^{-\nu t},
\end{equation*}

\noindent and we then observe that 
\begin{align*}
(\Delta - c_n R)(\vartheta e^{-\nu t}) &= -c_n \vartheta^\alpha \check{R} e^{-\nu t} + 2 \nabla \vartheta \cdot \nabla e^{-\nu t} + \vartheta \Delta e^{-\nu t} \\
&\leq (-c_n (\inf \vartheta)^\alpha \check{R}_0 + \nu A) e^{-\nu t}.
\end{align*}

\noindent From this it follows that there is some $\nu_0 > 0$ such that if $\nu \leq \nu_0$, there is a constant $k_\nu$ such that $(\Delta - c_n R)(\vartheta e^{-\nu t}) < -k_\nu e^{-\nu t}$. For any $\phi \leq \mr{\phi} + b \vartheta e^{-\nu t}$ with $\nu$ sufficiently small, we thus have that $\Lich_\phi (\mr{\phi} + b \vartheta e^{-\nu t})$ is bounded above by 
\begin{equation*}
(\Delta - c_n R)\mr{\phi} - b k_\nu e^{-\nu t} - \beta (\mr{\phi} + b \vartheta e^{-\nu t})^\alpha + c_n |\sigma + \DD W|^2 (\mr{\phi} + b \vartheta e^{-\nu t})^{-\gamma}.
\end{equation*}

\noindent One then argues exactly as in the proof of Proposition 3.5 to prove the result. \qed

\vspace{.5cm}

\noindent Just as in the AC case, we may choose $\nu$ sufficiently small and $b$ sufficiently large so that $\min\{\epsilon \vartheta, \mr{\phi} + b \vartheta e^{-\nu t}\}$ is a weak global supersolution.

The proofs of the previous two propositions suggest how we might construct a global subsolution. 

\begin{prop}
Let $\nu < \delta/(2\gamma - 2)$ be so small that $(\Delta - c_n R)(\vartheta e^{-\nu t}) < -k_\nu e^{-\nu t}$, and let $\phi_+$ be defined as above. Then there is some $a > 0$ such that if $\mr{\phi} - a \vartheta e^{-\nu t} \leq \phi \leq \phi_+$, we have that 
\begin{equation} \label{eq:lichineq}
\Lich_\phi (\mr{\phi} - a \vartheta e^{-\nu t}) \geq 0
\end{equation}
\noindent wherever $\mr{\phi} - a \vartheta e^{-\nu t} > 0$.
\end{prop}

\proof Carrying out computations nearly identical to those in section 3, we find that
\begin{equation*}
\Lich_\phi (\mr{\phi} - a \vartheta e^{-\nu t}) \geq \Lich_0 (\mr{\phi}) + a k_\nu e^{-\nu t} - c_n |\DD W|(|\DD W| + 2 |\sigma|)(\inf \mr{\phi})^{-\gamma}.
\end{equation*}

\noindent Since $|\Lich_0 (\mr{\phi})| \leq C_1 e^{-\omega t}$ by construction, the upper bound $\phi \leq \phi_+$ and Lemma 3.3 imply that for some constant $C_2 > 0$ we have 
\begin{equation*}
\Lich_\phi (\mr{\phi} - a \vartheta e^{-\nu t}) \geq a k_\nu e^{-\nu t} - C_1 e^{-\omega t} - C_2 e^{-\delta t}.
\end{equation*}

\noindent Therefore, as $\nu < \delta < \omega$, we need only choose $a$ so large that $a k_\nu > C_1 + C_2$ to ensure that (\ref{eq:lichineq}) holds. \qed

\vspace{.5cm}

It remains to construct a global subsolution on the bounded region where $\mr{\phi} - a \vartheta e^{-\nu t} \leq 0$. Choose $T > 0$ so large that $\mr{\phi} - a \vartheta e^{-\nu t} > \frac{1}{2} \inf \mr{\phi}$ whenever $t \geq T$. Since $R$ is bounded below and $\beta \geq \beta_0 > 0$, we may find some constant $K$ such that $c_n R + K \beta > 0$ everywhere. Letting $\mathcal{K} = \{t \leq T \}$, we may thus solve the boundary value problem
\begin{equation*}
\begin{cases}
(\Delta - c_n R - K \beta)\eta = 0 \\
\eta|_{\partial \mathcal{K}} = \frac{1}{2} \min\{1,\inf \mr{\phi}\}.
\end{cases}
\end{equation*}

\noindent The maximum principle implies that $\eta < 1$, and the strong maximum principle implies that $\eta > 0$. One easily verifies that $\eta$ is a global subsolution on $\mathcal{K}$, and that if we extend $\eta$ to be identically zero outside of $\mathcal{K}$, we find that $\max\{\eta,\mr{\phi} - \vartheta e^{-\nu t}\}$ is a weak global supersolution.

\subsection{Modifications to the fixed point argument}

Having constructed a pair of global sub-/supersolutions with the correct asymptotic limits, we may apply a fixed point argument similar to that found in section 4 to find a solution to (\ref{eq:lcby}). For this we define the set $U$ and the maps $\mathcal{W}$ and $\mathcal{Q}$ just as in section 4. We note that $\Lich_0 (\mathcal{Q}(\sigma)) = 0$, and so we simply replace ``$\mr{\phi}$'' with ``$\mathcal{Q}(\sigma)$'' in the definition of the map $\mathcal{S}_\sigma (\pi)$. The advantage of this replacement is that the conformal covariance of the Lichnerowicz equation gives us the identity
\begin{align}
\mcs_\sigma (\pi) = \QQ(\sigma + \pi) - \QQ(\sigma) &= \theta \hat{\QQ}(\theta^{-2} (\sigma + \pi)) - \theta \hat{\QQ}(\theta^{-2} \sigma) \label{eq:id1} \\
&:= \theta \hat{\mcs}_{\theta^{-2} \sigma} (\theta^{-2} \pi) \label{eq:id2}
\end{align}

\noindent where $\hat{\QQ}$ is the solution operator for the Lichnerowicz equation with respect to the metric $\theta^{4/(n-2)} g$. 

The proof of Theorem 2.2 reads exactly as the proof of Theorem 2.1 once we have established the analog of Lemma 4.2. For this we use an implicit function theorem argument very similar to the proof of that lemma, but we must deal with the fact that we no longer have $R > 0$. We shall get around this difficulty using a trick employed by Maxwell in \cite{M}, which essentially boils down to the identity (\ref{eq:id1}).

\begin{lem}
The map $\mathcal{S}_\sigma : C^{0,\mu} _{-\delta} (S^2 _0 (M)) \rightarrow C^{2,\mu} _{-\nu}(M)$, defined with respect to a conformally AC metric, is continuous.
\end{lem}

\proof Given $\pi_0 \in C^{0,\mu} _{-\delta} (S^2 _0 (M))$, we set $\theta_0 = \QQ(\sigma + \pi_0)$. If we also set $\hat{\sigma} = \theta_0 ^{-2} \sigma$ and $\hat{\pi}_0 = \theta_0 ^{-2} \pi_0$, and define $\hat{\mcs}_{\hat{\sigma}}$ as in (\ref{eq:id2}), we have $\hat{\mcs}_{\hat{\sigma}} (\hat{\pi}_0) = 1 - \hat{\QQ}(\hat{\sigma})$. We next define the map $F : C^{2,\mu} _{-\nu}(M) \times C^{0,\mu} _{-\delta}(S^2 _0 (M)) \rightarrow C^{0,\mu} _{-\nu}(M)$ by 
\begin{equation*}
F(\psi,\pi) = (\hat{\Delta} - c_n \hat{R})(\hat{\QQ}(\hat{\sigma}) + \psi) - \beta (\hat{\QQ}(\hat{\sigma}) + \psi)^\alpha + c_n |\sigma + \pi|^2 (\hat{\QQ}(\hat{\sigma}) + \psi)^{-\gamma}.
\end{equation*}

\noindent The Fr\'{e}chet derivative is computed exactly as in the proof of Lemma 4.2, and we find that at the point $(\hat{\psi}_0, \hat{\pi}_0) = (\hat{\mcs}_{\hat{\sigma}}(\hat{\pi}_0),\hat{\pi}_0) = (1 - \hat{\QQ}(\hat{\sigma}),\hat{\pi}_0)$, we have
\begin{equation*}
DF_{(\hat{\psi}_0,\hat{\pi}_0)} (h,0) = [ \hat{\Delta} - (c_n \hat{R} + \alpha \beta + c_n \gamma |\hat{\sigma} + \hat{\pi}_0|^2)]h.
\end{equation*}

\noindent However, as $\hat{\QQ}(\hat{\sigma}+\hat{\pi}_0) = 1$, we have $-c_n \hat{R} = \beta - c_n |\hat{\sigma}+\hat{\pi}_0|^2$, so that in fact
\begin{equation*}
DF_{(\hat{\psi}_0,\hat{\pi}_0)} (h,0) = [ \hat{\Delta} - ((\alpha - 1) \beta + c_n (\gamma +1) |\hat{\sigma} + \hat{\pi}_0|^2)]h.
\end{equation*}

\noindent The map $DF_{(\hat{\psi}_0,\pi)} : C^{2,\mu} _{-\nu} (M) \rightarrow C^{0,\mu} _{-\nu} (M) $ is thus an isomorphism, and so we conclude by the implicit function theorem that $\hat{\mcs}_{\hat{\sigma}}$ is continuous. The lemma now follows from (\ref{eq:id1}) and (\ref{eq:id2}), since we have $\mcs_\sigma (\pi) = \theta_0 \hat{\mcs}_{\hat{\sigma}} (\theta_0 ^{-2} \pi)$ for all $\pi \in C^{0,\mu} _{-\delta} (S^2 _0 (M))$. \qed

\vspace{.5cm}

As remarked above, the proof of Theorem 2.2 now goes through exactly as the proof of Theorem 2.1. We note that this theorem may be extended to handle the conformally AP case so long as the period of the metric on each end is a rational multiple of the periods of each vector field in $\mathscr{Y}_i$. This restriction would allow us to apply the quotient argument used to show that each vector field in $\mathscr{Y}_i$ is a conformal Killing field. Without this assumption, more work needs to be done to assure the solvability of (\ref{eq:lcby}) for the general conformally AP case.


\subsection*{Acknowledgement}

The author wishes to thank Rafe Mazzeo for his indispensable advice and helpful discussions throughout this paper's development.

\end{document}